\documentclass[letterpaper]{article}
\def\ket#1{|{#1}\rangle}
\def\bra#1{\langle{#1}|}

\def\i{\mathbf {i}}

\def\tr{\mathord{\mbox{\bf tr}}}

\usepackage{aaai}
\usepackage{times}
\usepackage{helvet}
\usepackage{courier}
\usepackage{pictex}
\begin{document}
%
\title{Certainty and Uncertainty in Quantum Information Processing}
\author{Eleanor G. Rieffel\\
FX Palo Alto Laboratory\\
rieffel@fxpal.com\\
}
\maketitle
\begin{abstract}
\begin{quote}

This survey, aimed at information processing researchers,
highlights intriguing but lesser known results, corrects
misconceptions, and suggests research areas.
Themes include: certainty in quantum algorithms; the ``fewer worlds"
theory of quantum mechanics; quantum learning; probability theory 
versus quantum mechanics.
\end{quote}
\end{abstract}


This idiosyncratic survey delves into areas of 
quantum information processing 
of interest to researchers in fields 
like information retrieval, machine learning, and artificial
intelligence. It overviews intriguing but
lesser known results, corrects common misconceptions,
and suggests research directions. Three types of applications
of a quantum viewpoint on information processing are discussed:
quantum algorithms and protocols; quantum proofs for classical results; the
use of formalisms developed for quantum mechanics in other areas
with linear algebraic or probabilistic components.
This paper is not tutorial in nature; readers new to the field 
should read it in conjunction with a 
tutorial \cite{Rieffel-00} or book \cite{NCbook,RPbook} on the subject.

A number of themes underlie this paper: certainty in quantum
algorithms and quantum mechanics, including a ``fewer worlds"
correction to popular conceptions of the ``many worlds" interpretation
of quantum mechanics; 
relations and distinct differences between
probability theory and quantum mechanics, including 
how entanglement differs from correlation; 
what is known and what remains uncertain
as to the source of the power of quantum information processing.
The most startling thing about quantum
mechanics is not that it is probabilistic, but rather that it
disobeys fundamental laws of probability theory. 
A common framework encompassing both probability theory and
quantum mechanics throws light on many
of these themes. The most technical parts of the paper 
establish this framework and discuss its implications.

\section{What is and isn't quantum information processing}

Quantum information processing includes quantum computation and
cryptographic and communication protocols like
quantum key distribution and dense coding. Quantum computation 
is not synonymous with using quantum effects in computation;
quantum mechanical effects are used in the processors of all state of
the art (classical) computers. The distinction between classical and
quantum computation is whether the information being processed
is encoded in a classical or quantum way, in bits or qubits.

\section{Certainty in quantum mechanics}

\subsection{Non-probabilistic quantum algorithms}

Glaringly obvious - perhaps blindingly so - examples of 
non-probabilistic quantum algorithms exist: quantum
analogs of classical non-probabilistic algorithms.
Any reversible classical computation has a directly analogous quantum
computation. Any classical computation has a reversible
counterpart using at most $O(t^{1+\epsilon})$ time and
$O(s\log t)$ space \cite{Bennett-89}.
If the initial classical algorithm is non-probabilistic,
so are the analogous reversible and quantum algorithms.

More surprising perhaps is that the first truly quantum algorithms 
- ones that do not have classical counterparts - succeed with 
certainty. The quantum algorithm for Deutsch's problem 
\cite{Deutsch-85,Deutsch-Jozsa-91} succeeds with certainty.
Grover's search algorithm is not inherently probabilistic. His
initial algorithm succeeded only with high probability \cite{Grover-97},
but with a little cleverness Grover's algorithm
can be modified so that it is guaranteed to find an
element being searched for while still preserving the quadratic
speed up.  
\cite{Brassard-Hoyer-Tapp} suggest two approaches. In essence, the first 
rotates by a slightly smaller angle at each step, while
the second changes only the last step to a smaller rotation.
Shor's factoring algorithm is inherently probabilistic just like 
many of the best classical algorithms for related problems like
primality testing.

\subsection{Fewer worlds theory of quantum mechanics}

Many papers discuss the pros and cons of the many worlds theory. 
Here we mean to correct not that theory, but the popular conception
of it as ``everything happens in some universe". 
Popular accounts of quantum mechanics, and some scholarly articles,
give the impression that quantum mechanics, at least in the many
worlds interpretation, implies that everything happens in some universe.
A typical quote \cite{Deutsch-98}: 
``There are even universes in which a given object in our universe 
has no counterpart - including universes in which I was never born 
and you wrote this article instead." 
The variety of imaginative examples suggest that
anything we can conceive of, even the highly unlikely,
happen, if only in a small number of universes. 
But much of the surprise of quantum mechanics is that certain things
we thought would happen, even things we thought were sure to happen,
do not happen at all. 

Most startling are events that were predicted
to happen with certainty by classical physics, but which in fact happen
with probability $0$. Thus, not only is it not true that everything we
can conceive of is predicted to happen in some 
universe, but things we can hardly conceive of not happening do not
happen, not in any universe.
To emphasize this correction, I call it ``the fewer worlds than
we might think" interpretation of quantum mechanics, or the
``fewer worlds" theory for short.

Here are a few examples. In
the double slit experiment, quantum mechanics predicts that no
light reaches certain spots. And indeed no light reaches those spots, 
even though classically we expect some photons to reach
every spot. Even more striking is the GHZ experiment 
\cite{GHZ-89,GHZ-90,Pan-00} in which the classical
prediction is that each of four things happen with equal probability
and another four things never happen. Quantum mechanics predicts, 
and experiments confirm, that the four outcomes that are classically
predicted to happen never happen (and the four classically prohibited
outcomes do occur, with equal probability). 
As a final example, we saw that many quantum algorithms 
return a result with probability $1$; the obvious conclusion is that
the other results do not happen at all.

\section{Uncertainty in classical physics}

Both relativity and uncertainty principles exist in purely classical
settings. The revolutions of the $20^{th}$ century, special and general
relativity and quantum mechanics, expanded on these principles.
In special relativity, Einstein made Galilean relativity - the notion
that the speed of an object depends on the observer and is not a
property of the object itself - compatible with the notion of a
constant speed of light, the same for all observers. Quantum mechanics
took standard classical uncertainty principles involving waves
and applied them to particles with the implication that nothing
of a pure particle nature exists, in this way resolving various
experimental and theoretical issues.

That a particle cannot simultaneously have both a precisely defined 
position and a precisely defined momentum is the startling content of
Heisenberg's uncertainty principle. This statement is less
surprising when applied to a wave. Uncertainty principles for
classical waves are well known. For example, consider a signal $s(t)$
with a finite mean $\hat{t}$ and standard deviation $\Delta t$.
Similarly assume the mean $\hat\omega$ and standard deviation 
$\Delta \omega$ of $s(t)$'s frequency distribution can be calculated. 
Classical signals $s(t)$ obey the uncertainty principle 
$\Delta t \Delta \omega \geq 1/2$. That a signal with small standard
deviation in time cannot have too small a standard deviation in its
frequency spectrum is not mysterious. Details can be found
in many signal processing books; \cite{Cohen-95} is particularly 
detailed and insightful.

This discussion makes no mention
of measurement (though it certainly has implications for measurement).
Contrary to popular belief, Heisenberg's uncertainty principle is
not about imprecision in our ability to measure 
(though it has implications for measurement). Just like time/frequency
in the signal case, 
Heisenberg's uncertainty principle says that a particle cannot
have definite values for both its position and momentum. 
The implication is that there are no classical point particles, with 
position and momentum both precisely defined; there aren't even 
arbitrary close approximations to such.
The implications of this principle for measurement is that even
in an ideal case, in which measurement of a series of particles in
identical states were performed perfectly,
if the standard deviation of the results for position measurements
is small enough then the standard deviation of the results for
momentum must be proportionally large.
Initially Heisenberg and others confused two arguments, one based on the
wave nature of particles, the other based on a disturbance theory
of measurement. It is the former that has stood 
the test of time. The failure of a disturbance theory was established 
by the famous EPR paper \cite{EPR} (though it took decades before
a fuller understanding of the implications of the EPR paradox 
was achieved by Bell).

Generalized uncertainty principles exist for many other pairs
of properties. For example, an uncertainty relation for
polarization says that if a particle has polarization close
to horizontal or vertical it cannot have polarization close to
$45^\circ$. This uncertainty principle is more intuitive than
that for position and momentum, but the mathematics is closely
related. 

\section{Applications of a quantum viewpoint to information processing}

There exist three distinct classes of applications of the viewpoint 
that has developed from the study of quantum information processing.
The first and most obvious class contains quantum
algorithms and protocols. The second is the use of reasoning about
quantum systems to obtain insight into classical computer science. 
The third class consists of purely classical results inspired by
the formalisms developed to deal with quantum information processing
and quantum mechanics more generally. We briefly discuss this last
class of applications, and then devote a section to each of the
first two classes.

Researchers in quantum mechanics, responding to their need to delve
deeply and carefully into the linear algebra and generalized
probability theory underlying quantum mechanics, have developed
powerful formalisms for discussing these areas. 
Dirac's compact and suggestive bra/ket notation is useful for any
work involving significant linear algebra. 
The operator view gives insight into classical probability 
theory, and understanding the tensor structure inherent in classical 
probability theory and its difference from a direct sum structure
helps clarify many issues including relationships between joint
distributions and their marginals.

\subsection{Implications of reasoning about quantum systems to 
problems in classical computer science}

We give two surprising, elegant examples.

Cryptographic protocols usually rely on the empirical hardness of
a problem for their security; it is rare to be able to prove
complete, information theoretic security. When a cryptographic protocol 
is designed based on a new problem, the 
difficulty of the problem must be established before the security
of the protocol can be understood. Empirical testing of a problem
takes a long time. Instead, whenever possible, 
``reduction" proofs are given that show that if the new problem were
solved it would imply a solution to a known hard problem; 
the proofs show that the solution to the known problem can be reduced
to a solution of the new problem. 
\cite{Regev-05} designed a novel, purely classical cryptographic
system based on a certain problem. He was able to reduce a known hard
problem to this problem, but only by using a quantum step
as part of the reduction proof. Thus he has shown that if the new
problem is efficiently solvable in any way, there is an efficient
quantum algorithm for the old problem. But it says nothing about whether
there would be a classical algorithm. This result is of practical 
importance; his new cryptographic algorithm is a more efficient lattice based
public key encryption system.  Lattice based systems are
currently the leading candidate for public key systems secure against
quantum attacks.

More spectacular, if less practical, is Aaronson's new solution to
a notorious conjecture involving a purely classical complexity class
{\bf PP} \cite{Aaronson04}.
From 1972 until 1995 this question remained open.  Aaronson
defines a new quantum complexity class {\bf PostBQP}, an extension of the
standard quantum complexity class {\bf BQP}, motivated by the use of
postselection in certain quantum arguments. It takes him a page to show
that {\bf PostBQP}={\bf PP}, and then only three lines to prove the
conjecture. Thus it seems that for certain questions, the ``right"
way to view the classical class {\bf PP} is through the eyes of
quantum information processing.

\subsection{Quantum algorithms and protocols}

Shor's factoring and discrete log algorithms solve important but
narrow problems. Grover's algorithm and its generalizations are
applicable only to a more restricted class of problems than many
people outside the field realize. For example, it is unfortunate
that Grover used ``database" in the title of 
\cite{Grover-97} since his algorithm does not apply to what most
people mean by a database. Grover's algorithm only gives a 
speed-up over unstructured search, and databases, which are
generally highly structured, can be searched extremely rapidly
classically. At best
quantum computation can only give a constant factor improvement
for searches of ordered data like that of databases
\cite{Childs-06}.

Even worse, obtaining output from Grover's algorithm destroys the 
quantum superposition, and recreating the
superposition is often linear in $N$
which negates the $O(\sqrt N)$ benefit of the search algorithm.
For this reason Grover's algorithm and its generalizations are only 
applicable to searches over data that has a sufficiently uniform
and quick generating function which can be used to quickly 
compute the superposition.

Finding new quantum algorithms has been exceedingly slow going. Some
more recent algorithms include \cite{Hallgren-02} for solving Pell's 
equations, \cite{Watrous-01} for the group black box model,
\cite{vanDam-03} for the shifted Legendre symbol problem. 
The first two are closely related to Shor's algorithm - they are in the
class of hidden subgroup problems - and the third makes heavy use
of Fourier transforms. 
In the past five years a new family of quantum algorithms
has been discovered that uses techniques of quantum walks 
to solve a variety of problems, some related to graphs, others
to matrix products or commutativity in groups
\cite{Childs-02,Magniez-05a,Magniez-05b,Buhrman-06,Krovi-07}.

For many years Shor's algorithm and Grover's algorithm were viewed
as widely different algorithms. Quantum learning theory 
\cite{Bshouty-99,Servedio-01,Gortler-04,Hunziker-03,Atici-04}
is closely tied to both. Quantum learning descends from
computational learning theory, a subfield of artificial intelligence.
Computational learning is concerned with concept
learning. 
Common models include exact learning and probably approximately correct 
(PAC) learning.
A concept is modeled by its membership as given
by a Boolean function $c:\{0,1\}^n \to \{0,1\}$. Let $C = \{c_i\}$ be
a concept class. Say one has access to an oracle $O_{c}$
for one of the concepts $c$ in $C$, but one doesn't know which. The
types of oracles assumed vary, but a common one is a membership
oracle which upon input of $x$ outputs $c(x)$. 
In the quantum case, one can input superpositions of inputs to
obtain superpositions of outputs. 
One can ask a variety of questions as to how quickly and with
how many queries to the oracle can the concept $c$ be determined.
Sample results in this area
include the negative result that the number of classical and
quantum queries required for any concept class does not differ
by more than a polynomial in either the exact or PAC model. However
the story is different if computational efficiency is taken into account. 
In the exact model the existence of {\em any} classical one-way
function guarantees the existence of a concept class which is 
polynomial-time learnable in the quantum case but not in the classical.
For the PAC model a slightly weaker result is known in terms of
a particular one-way function.

\section{Probability theory and quantum mechanics}

To quote Scott Aaronson \cite{Aaronson05}:

``To describe a state of n particles, we need to write down an 
exponentially long vector of exponentially small
numbers, which themselves vary continuously. Moreover, the instant 
we measure a particle, we ``collapse" the vector that describes 
its state - and not only that, but possibly the state of another 
particle on the opposite side of the universe. Quick, what theory
have I just described?''

``The answer is classical probability theory. The moral is that, before 
we throw up our hands over the ``extravagance" of the quantum worldview, 
we ought to ask: is it so much more extravagant than the classical
probabilistic worldview? After all, both involve linear 
transformations of exponentially long vectors that
are not directly observable."

We spend the next section putting this view of probability theory
on a firm basis. We then describe how quantum
mechanics is a formal extension of probability theory. 
We only sketchily describe this extension; more
details can be found in 
\cite{Strocchi,Kuperberg,Summers-06,Kitaev-02,Sudbery-86,Mackey}.

Many, but not all, of the unintuitive aspects of quantum
mechanics exist in classical probability theory. Entanglement does
not exist in classical probability, but classical correlations are
strange enough, judging by human reaction to many of them.

\subsection{A view of classical probability theory}

Let $A$ be a set of $n$ elements. A probability distribution $\mu$
on $A$ is a function 
$$\mu: A\to [0,1]$$
such that $\sum_{a\in A} \mu(a) = 1$.
The space ${\cal P}^A$ of all probability distributions over $A$ has dimension
$n-1$. We can view  ${\cal P}^A$ as the $n-1$ dimensional
simplex $\sigma_{n-1} = \{x\in{\bf R}^n | x_i \geq 0, x_1 +x_2 + \cdots + x_n =1 \}$
which is contained in the $n$ dimensional space ${\bf R}^A$, the space
of all functions from $A$ to ${\bf R}$,
$${\bf R}^A = \{f: A \to {\bf R} \}.$$
For $n = 2$, the simplex $\sigma_{n-1}$ is the line segment from
$(1,0)$ to $(0,1)$.
The vertices of the simplex correspond to the elements $a\in A$: a probability
distribution $\mu$ maps to the point in the simplex 
$x = (\mu(a_1), \mu(a_2), \dots, \mu(a_n) )$.

Let $B$ be a set of $m$ elements. Let $A\times B$ be the Cartesian
product $A\times B = \{(a,b)| a\in A, b\in B \}$.
What is the relation between ${\cal P}^{A\times B}$, the space of all
probability distributions over $A\times B$, and the spaces
${\cal P}^A$ and ${\cal P}^B$? 
The tempting guess is not correct: 
${\cal P}^{A\times B} \ne {\cal P}^A \times{\cal P}^B$.
We see this relation does not hold by checking
dimensions.  First consider the relationship between
${\bf R}^{A\times B}$ and ${\bf R}^A$ and ${\bf R}^B$.
Since $A\times B$ has cardinality $|A\times B| = |A||B| = nm$,
${\bf R}^{A\times B}$ has dimension $nm$, which is not equal to
$n+m$, the dimension of  ${\bf R}^A \times {\bf R}^B$.
Since in general $dim({\cal P}^A) = dim({\bf R}^A) - 1$, 
$dim({\cal P}^{A\times B}) = nm -1$ which is not equal to $n + m -2$, 
the dimension of ${\cal P}^A \times {\cal P}^B$, so
${\cal P}^{A\times B} \ne {\cal P}^A \times {\cal P}^B$.
Instead ${\bf R}^{A\times B}$ is the tensor product 
${\bf R}^A \otimes {\bf R}^B$ of
${\bf R}^A$ and ${\bf R}^B$. 
So ${\cal P}^{A\times B}\in {\bf R}^A \otimes {\bf R}^B$.

Tensor products are rarely mentioned in probability textbooks, but
the tensor product is as much a part of probability theory as of
quantum mechanics. The tensor product structure inherent in probability
theory should be stressed more often;
one of the sources of mistaken
intuition about probabilities is a tendency to try to impose the
more familiar direct product structure on what is actually a 
tensor product structure.
We briefly review tensor products here; readers not familiar
with tensor products should consult more extensive expositions
\cite{Rieffel-00,NCbook,RPbook}.

The {\it tensor product} $V \otimes W$
of two vector spaces $V$ and $W$ with bases
$A = \{{\bf a}_1, {\bf a}_2, \dots, {\bf a}_n\}$ and
$B = \{{\bf b}_1, {\bf b}_2, \dots, {\bf b}_m\}$ respectively is
an $nm$-dimensional vector space with basis
${\bf a}_i\otimes{\bf b}_j$ where
$\otimes$ is the tensor product, an abstract binary operator 
defined by the following relations:
\begin{eqnarray*}
({\bf v}_1 + {\bf v}_2)\otimes {\bf x}&=& {\bf v}_1\otimes{\bf x} + {\bf v}_2 \otimes {\bf x}\\
{\bf v}\otimes({\bf x}_1+ {\bf x}_2)&=& {\bf v}\otimes {\bf x}_1+ {\bf v} 
\otimes {\bf x}_2\\ (\alpha {\bf v}) \otimes {\bf x} &=& {\bf v} \otimes (\alpha {\bf x}) = \alpha \ {\bf v} \otimes {\bf x}.
\end{eqnarray*}
Taking $k = \min(n,m)$,
all elements of $V \otimes X$ have form
$${\bf v}_1\otimes{\bf w}_1 +
{\bf v}_2\otimes{\bf w}_2  + \cdots +
{\bf v}_k\otimes{\bf w}_k.$$
Due to the relations defining the tensor product such
a representation is not unique. Furthermore,
most elements of $V\otimes W$
cannot be written as $v\otimes w$ where $v\in V$ and $w\in W$.

Let $A_0 = \{0_0, 1_0\}$, $A_1 = \{0_1, 1_1\}$, and $A_2 = \{0_2, 1_2\}$,
where $1_0$ versus $0_0$ corresponds to whether or not the
next person you meet is interested in quantum mechanics, $A_1$ to whether 
they know the solution to the Monty Hall problem, and $A_2$
to whether they are at least $5'6''$ tall. So $1_0 1_0 0_0$ corresponds to 
someone under $5'6''$ who is interested in quantum mechanics and knows
the solution to the Monty Hall problem. We often write $110$ instead
of $1_0 1_0 0_0$; the subscripts are implied by the position. A
probability distribution over the set of eight possibilities, 
$A_0\times A_1\times A_2$, has form
$$\vec{p} = 
(p_{000}, p_{001}, p_{010}, p_{011}, p_{100}, p_{101}, p_{110}, p_{111}).$$
More generally, a probability distribution over 
$A_0\times A_1\times \cdots\times A_k$, where the $A_i$ are all
$2$ element sets, is a vector of length $2^k$. We now understand
the first part of Aaronson's remark: vectors in probability
theory are exponentially long.

Given functions $f: A \to {\bf R}$ and $g: B\to {\bf R}$, define the tensor
product $f\otimes g: A\times B \to {\bf R}$ by
$(a, b)\mapsto f(a)g(b)$. 
If $\mu$ and $\nu$ are probability distributions,
then so is $\mu\otimes \nu$.
The linear combination of distributions is a distribution as long 
as the linear coefficients are non-negative and sum to 1.
Conversely, any distribution 
$\eta\in {\cal P}^{A\times B}$ is a linear combination of
distributions of the form $\mu\otimes \nu$
with linear factors summing to $1$.

A joint distribution $\mu\in {\cal P}^{A\times B}$ is 
{\it independent} or {\it uncorrelated} if it can be written as 
a tensor product $\mu_A\otimes \mu_B$ of distributions
$\mu_A\in {\cal P}^A$ and $\mu_B\in {\cal P}^B$.
The vast majority of
joint distributions do not have this form, in which case they are 
{\it correlated}.
For any joint distribution
$\mu\in {\cal P}^{A\times B}$, we can define
a {\it marginal}
distribution $\mu_A\in {\cal P}^A$ by
$$\mu_A: a\mapsto \sum_{b\in B} \mu(a,b).$$
An uncorrelated distribution is the tensor product of
its marginals. Other distributions cannot
be reconstructed from their marginals; information has been lost. 

A distribution $\mu$ on a finite set $A$ that is concentrated entirely
at one element is said to be a {\it pure}; on a
set $A$ of $n$ elements there are exactly $n$ pure distributions
$\mu_a: A\to [0,1]$, one for each element of $A$, where
$$
\mu_a: a'\mapsto \left\{
\begin{array}{cc}
1 & \mbox{if } a' = a \\
0 & \mbox{otherwise}.
\end{array} \right.$$
All other distributions are said to be 
{\it mixed}\index{mixed distribution}\index{distribution ! mixed}.

Let us return to the example of the traits for the next person
you meet. Unless you know all of these traits, the
distribution $\vec{p} = (p_{000},\dots , p_{111})$
is a mixed distribution.
When you meet the person you can observe their traits.
Once you have made these
observations, the distribution ``collapses" to a pure 
distribution. For example, if the person is interested in quantum
mechanics, does not know the solution to the Monty Hall problem, and
is $5'8''$, the ``collapsed" distribution is
$\vec{p_{c}} = (0,0,0,0,0,1,0,0).$

To understand the final part of Aaronson's remark, consider another
example. Say someone prepares two sealed envelopes with identical 
pieces of paper and sends them to opposite sides of the universe. 
Half the time both envelopes contain $0$; half the time $1$. 
The initial distribution is
$\vec{p_I} = (1/2,0,0,1/2).$
If someone then opens one of the envelopes and observes a $0$, the
state of the contents of the other envelope is immediately known - 
known faster than light can travel between the envelopes - and the
distribution ``collapses" to 
$\vec{p_u} = (1,0,0,0).$

Are we disturbed by the ``extravagance" of the exponential 
state space of classical probability theory, and the 
``faster-than-light collapse" of these classical vectors under observation?
Another question one might ask is: can this ``extravagance" be used to
facilitate computation? The answer is a resounding yes; allowing
randomness does give additional computational power. See 
\cite{Harel-87,Traub-99} for delightful expositions of
the computational benefits of randomness.

To fully understand the relationship between quantum mechanics and
probability theory it is useful to view probability distributions as
operators. Consider the set of linear operators
${\cal M}^A = \{M: {\bf R}^A \to {\bf R}^A \}$. To every function
$f: A\to {\bf R}$, there is an associated operator 
$M_f: {\bf R}^A\to {\bf R}^A$ given by
$M_f: g\mapsto fg$.  An operator $M$ is said to be a projector 
if $M^2 = M$. The probability distributions $\mu$ whose corresponding
operators $M_\mu$ are projectors are exactly the pure distributions.
The matrix for the operator corresponding to a function is always
diagonal; for a probability distribution, diagonal and trace $1$.
For example, the operator corresponding to the probability
distribution $\vec{p_I} = (1/2,0,0,1/2)$ is represented by the matrix
\begin{eqnarray*}
\left(
\begin{array}{cccc}
1/2 & 0 & 0 & 0 \\
0 & 0 & 0 & 0 \\
0 & 0 & 0 & 0 \\
0 & 0 & 0 & 1/2\\
\end{array}
\right).
\end{eqnarray*}

\subsection{Quantum mechanics as a generalization of probability theory}

The vector representation of a quantum state has redundancy that 
can be confusing; 
any vector multiplied by a unit length complex number $e^{\i\theta}$
- called the global phase - represents the same quantum state.
Another way of representing quantum states 
removes this ambiguity and makes the relation with probability
theory clearer.
We follow Dirac's elegant and compact bra/ket notation.
The row vector $\bra v$ is the conjugate transpose of the column vector
$\ket v$. 
For any $N$ dimensional vector $\ket v$ representing a quantum
state we can construct a density
operator, the $N\times N$ matrix $\ket v\bra v$. 
The density operator $\ket v\bra v$ representing
a quantum state no longer has ambiguity due to the global phase.
Like the operators corresponding to probability distributions, the
operators corresponding to quantum states have trace $1$ and are
positive and Hermitian.
Density operators corresponding to quantum states
$\ket v$ are projectors so have rank $1$.
Unlike operators for probability distributions,
density operators need not be diagonal.
For example, the density operator for the state 
$\ket\nearrow = 1/\sqrt{2}(\ket 0 + \ket 1)$ is
\begin{eqnarray*}
\ket\nearrow\bra\nearrow = \left(
\begin{array}{cc}
1/2 & 1/2 \\
1/2 & 1/2 \\
\end{array}
\right)
\end{eqnarray*}
This example illustrates that superpositions are distinct from
mixtures of basis states 
since such mixtures must be diagonal: the fifty-fifty mixture of 
$\ket 0$ and $\ket 1$ has density operator 
\begin{eqnarray*}
\frac{1}{2}(\ket 0\bra 0 + \ket 1\bra 1) = \left(
\begin{array}{cc}
1/2 & 0 \\
0 & 1/2 \\
\end{array}
\right)
\end{eqnarray*}

The analog of taking the marginal is taking the partial trace. 
The partial trace $\tr_W O_{VW}$ of an operator 
$\rho: V\otimes W \to V\otimes W$ with respect to the 
subsystem $W$ is the operator
$$\rho_V = \tr_W O_{VW} = \sum_i \bra{b_i} O_{VW} \ket{b_i}$$
that acts on subsystem $V$,
where $\{\ket{b_i}\}$ is a orthonormal basis for $W$.
Taking the partial trace of a density operator produces another
density operator, a Hermitian, positive, trace $1$ operator.
Density operators obtained from the partial trace model what
can be learned about a subsystem from measurements on that
subsystem alone. In this context they are often called mixed
states. Density operators of the form $\ket v\bra v$ are called
pure states, or just quantum states.
For example, the Bell state 
$\ket{\Phi^+} = 1/\sqrt 2(\ket{0}\otimes\ket{0} + \ket{1}\otimes\ket{1})
= 1/\sqrt 2(\ket{00} + \ket{11})$
has density operator
$$\ket{\Phi^+}\bra{\Phi^+} = 
\frac{1}{2}\left(\begin{array}{cccc}
1 & 0 & 0 & 1 \\
0 & 0 & 0 & 0 \\
0 & 0 & 0 & 0 \\
1 & 0 & 0 & 1 \\
\end{array}\right),$$
and its partial trace with respect to either one of its qubits
is the $2$-dim density operator $\frac{1}{2} I$.

Since every Hermitian operator can be diagonalized, every density 
operator $\rho$ can be written as $\sum_i p_i \ket{\psi_i}\bra{\psi_i}$,
a probability distribution over pure quantum states 
where the $\ket{\psi_i}$ are mutually orthogonal eigenvectors 
of $\rho$, and $p_i$ are the eigenvalues.
Conversely any probability
distribution $\mu$ over a set of orthogonal quantum states
$\ket{\psi_1}, \ket{\psi_2}, \dots, \ket{\psi_L}$ where 
$\mu:\ket{\psi_i} \to p_i$ has a corresponding density operator
$\rho_\mu = \sum_i p_i \ket{\psi_i}\bra{\psi_i}$. In the basis 
$\{ \ket{\psi_i} \}$, the density operator $\rho_\mu$ is diagonal
with entries $p_1, \dots, p_L$.
Under the isomorphism between ${\bf R}^L$ and the subspace
of $V$ generated by $\ket{\psi_1}, \ket{\psi_2}, \dots, \ket{\psi_L}$,
the density operator $\rho_\mu$ realizes the operator $M_\mu$.
Thus a probability distribution
over a set of orthonormal quantum states $\{ \ket{\psi_i} \}$ can be 
viewed as a trace $1$ diagonal matrix acting on ${\bf R}^L$.

Although every density operator can be viewed as a probability 
distribution over a set of orthogonal quantum states, this
representation is not in general unique. 
More importantly, for most pairs of density operators $\rho_1$ and
$\rho_2$, there is no basis over which both $\rho_1$ and $\rho_2$
are diagonal. In particular, only if $\rho_1$ and $\rho_2$ commute are
they simultaneously diagonalizable, so only in this case can they both
be viewed as probability distributions over the same set of states.
Thus, although each density operator of dimension $N$ can
be viewed as a probability distribution over $N$ states, the space
of all density operators is much larger than the space of probability
distributions over $N$ states. 
Let $\rho: V\to V$ be a density operator. 
A density operator $\rho$ corresponds to a pure
state if and only if it is a projector. This statement is analogous to
that for probability distributions;
the pure states correspond exactly to rank $1$ density
operators, and mixed states have rank greater than $1$.
Density operators are also used to model probability distributions
over pure states, particularly probability distributions over the
possible outcomes of a measurement yet to be performed. Their use
here is analogous to the classical use of probability distributions 
to model the probabilities of possible traits before they can be observed.

A pure quantum state $\ket\psi$ is {\it entangled} if it
cannot be written as the tensor product of single qubit states.
For a mixed
quantum state, it is important to determine if all of its correlation
comes from being a mixture in the classical sense or
if it is also correlated in a quantum fashion. A mixed quantum state 
$\rho:V\otimes W\to V\otimes W$
is said to be {\it uncorrelated}\index{uncorrelated} if
$\rho = \rho_V\otimes \rho_W$ for some density operators 
$\rho_V: V\to V$ and $\rho_W: W\to W$.
Otherwise $\rho$ is said to be {\it correlated}\index{correlated}.
A mixed quantum state $\rho$ is said to be
{\it separable} if it can be written
$\rho = \sum_{j=1}^L p_j \ket{\psi_j^V}\bra{\psi_j^V}\otimes 
\ket{\phi_j^W}\bra{\phi_j^W}$ where $\ket{\psi_j^V}\in V$
and $\ket{\psi_j^W}\in W$. In other words, $\rho$ is
separable if all the correlation comes from its being a classical
mixture of uncorrelated quantum states. If a mixed state $\rho$
is not separable it is {\it entangled}.
For example, the mixed state 
$\rho_{cc} = \frac{1}{2}(\ket{00}\bra{00}) + (\ket{11}\bra{11})$
is classically correlated but not entangled whereas the Bell state
$\ket{\Phi^+}\bra{\Phi^+} = \frac{1}{2}(\ket{00} +\ket{11})(\bra{00} +\bra{11})$
is entangled.
The marginals of a pure distribution are always pure, but the
analogous statement is not true for quantum states; all of
the partial traces of a pure state are pure only if the original 
pure state was not entangled. 
As we saw, the partial traces of the Bell state $\ket{\Phi^+}$,
a pure state, are not pure.  Most pure
quantum states are entangled, exhibiting quantum correlations with
no classical analog. All pure probability distributions 
are completely uncorrelated. 

Classical and quantum analogs: 

\begin{tabular}{c|c}
   Classical probability& Quantum mechanics \\ \hline\hline
probability distribution & density operator $\rho$\\ 
$\mu$ viewed as operator $M_\mu$ & \\ \hline
pure dist: & pure state: \\
$M_\mu$ is a projector & $\rho$ is a projector \\ \hline
marginal distribution & partial trace \\ \hline
A distribution is & A state is \\
{\it uncorrelated} & {\it uncorrelated} \\
if it is the tensor product & if it is the tensor product \\
of its marginals & of its partial traces \\ \hline
\end{tabular}

Key difference:

\begin{tabular}{c|c}
Classical probability & Quantum mechanics \\ \hline\hline
pure distributions are & pure states contain \\
always uncorrelated & no classical correlation \\
  & but can be entangled \\ \hline
\end{tabular}

\section{Where does the power of quantum information processing come from?}

\subsection{Quantum parallelism?}

For any classical computation of a function $f(x)$ on $n$ bits,
the analogous quantum computation $U_f$ 
produces a superposition $\frac{1}{\sqrt N}\sum \ket{x,f(x)}$
of all input/output pairs upon input
of a superposition of all input values.
The ability of a quantum computer to obtain a superposition
of all input/output pairs with similar effort as it takes a
classical computer to obtain a single pair is called
{\em quantum parallelism}.
Since quantum parallelism 
enables one to work simultaneously with $2^n$ values,
it in some sense circumvents the time/space trade-off
of classical parallelism through its ability to hold exponentially
many computed values in a linear amount of physical space. However,
this effect is less powerful than it may initially appear.

We can gain only limited information from this superposition:
these $2^n$ values of $f$ are not independently accessible. We 
only gain information by measuring, but
measuring in the standard basis projects the final state
onto a single input/output pair $\ket{x, f(x)}$, and a random one at
that. By itself, quantum parallelism is useless. 

While $N = 2^n$ output values of $f(x)$ appear
in the single superposition state, it still takes 
$N = 2^n$ computations of $U_f$
to obtain them all, no better than the classical case.
This limitation leaves open the possibility that quantum parallelism
can help in cases where only a single output, or a small number of outputs,
is desired. It suggests an exponential speed up, but such speed ups
are rare. It has been proven that no quantum algorithm can improve
on the $O(\sqrt{N})$ that Grover's algorithm achieves for unstructured
search \cite{Bennett-et-al-97}, and for many other problems it has been 
proven that quantum computation cannot provide any speed-up 
\cite{Beals-01,Ambainis-00}.

\subsection{Exponential size of quantum state space?}

As we have seen, exponential spaces also arise in classical probability 
theory. Furthermore, what would it mean for an efficient algorithm to
take advantage of the exponential size of a space?
A superposition like $\frac{1}{\sqrt N}\sum \ket{x,f(x)}$ is
only a single state of the quantum state space.
The vast majority of states cannot even be approximated by an 
efficient quantum algorithm \cite{Knill-95}.
An efficient quantum algorithm cannot even come close to most
states in the state space.
So quantum parallelism does not, and efficient quantum algorithms
cannot, make use of the full state space.  

\subsection{Quantum Fourier transforms?}

Most quantum algorithms use quantum Fourier transforms (QFTs). The
Walsh-Hadamard transformation, a QFT over
the group ${\bf Z}_2$, is frequently used
to create a superposition of $2^n$ input values.
In addition the heart of most quantum algorithms makes use of 
QFTs. Shor and Grover use QFTs in both of these ways. 
Many researchers speculated that quantum
Fourier transforms are the paramount quantum resource for
quantum computation. So it came as a surprise when 
\cite{Aharonov-06} showed that the QFT
is classically simulatable. Given the ubiquity of quantum
Fourier transforms in quantum algorithms, researchers continue
to consider QFTs as one of the main tools
of quantum computation, but in themselves they are not sufficient.

\subsection{Entanglement?}

\cite{Jozsa-Linden} show that any quantum algorithm involving only
pure states that achieves exponential speed-up over classical algorithms 
must entangle a large number of qubits. 
While entanglement is necessary for an exponential speed-up,
the existence of entanglement is far from sufficient to guarantee
a speed-up, and it may turn out that another property better
characterizes what gives a speed-up. Many entangled systems have
been shown to be classically simulatable
\cite{Vidal-03,Markov-05}. Furthermore, if one looks at query complexity 
instead of algorithmic complexity, an
exponential benefit can be obtained without any entanglement
whatsoever. \cite{Meyer-00} shows that in the course of the
Bernstein-Vazirani algorithm, which achieves an $N$ to $1$
reduction in the number of queries required, no qubits become
entangled. More obviously the BB84 quantum key distribution
protocol makes no use of entanglement.

For these reasons entanglement should not be viewed as the sole 
source of power in quantum information processing. However
it is important in many contexts, and required in others. While
researchers have long recognized entanglement as a uniquely
quantum resource, much about entanglement is poorly understood.
Entanglement with respect to tensor decompositions of only two
factors is completely characterized for pure states, and well
studied for mixed states. 
See \cite{Bruss} for an introductory survey.
But understanding bi-partite entanglement
is of limited utility for understanding quantum computation
because there we are interested in entanglement between large 
numbers of qubits. 
Full characterization of entanglement with
respect to tensor decompositions with many factors is difficult;
where in the bi- or tri-partite cases only a finite number of
parameters are needed, infinitely many parameters are required for
four or more tensor factors \cite{Dur}.

Instead of trying to fully characterize multipartite entanglement,
we can ask which types of entanglement are useful, and for what.
Significant progress has been made here, though much work remains.
Cluster states were discovered to be a universal resource for
quantum computation. In cluster state, or one-way, quantum computing
\cite{Raussendorf-03,Nielsen-05}
a highly entangled ``cluster" states is set up at the beginning
of the algorithm. All computations take place by single qubit
measurements, so the entanglement between the qubits can only decrease
in the course of the algorithm (the reason for the ``one-way" name).
The initial cluster state is independent of the
algorithm to be performed; it depends only on the size of the problem
to be solved. In this way cluster state quantum computation makes
a clean separation between the entanglement creation and the
computational stages. While the cluster state model clarifies somewhat
the role of entanglement in quantum computation, in another model, adiabatic
quantum computation \cite{Aharonov-04}, which like the cluster state 
model has been proved
equivalent to the standard circuit model of quantum computation, 
the role of entanglement is obscure. Many intriguing questions as to the 
source of power in quantum information processing remain, and
are likely to remain for many years while we humans struggle to
understand what Nature allows us to compute quickly and why.

\bibliography{qcCopy}
\bibliographystyle{aaai}
\end{document}